\documentclass[journal,twoside,web]{IEEEtran}
\usepackage{amsmath,amsfonts}
\usepackage{algorithmic}
\usepackage{array}
\usepackage[caption=false,font=normalsize,labelfont=sf,textfont=sf]{subfig}
\usepackage{textcomp}
\usepackage{stfloats}
\usepackage{url}
\usepackage{verbatim}
\usepackage{graphicx}
\usepackage{algorithm}
\usepackage{setspace}
\usepackage{color}
\usepackage{cite}
\usepackage{booktabs}
\newtheorem{remark}{\bf \ \ Remark}

\newtheorem{theorem}{\bf \ \ Theorem}
\newtheorem{lemma}{\bf \ \ Lemma}
\newtheorem{assumption}{\bf \ \ Assumption}

\def\BibTeX{{\rm B\kern-.05em{\sc i\kern-.025em b}\kern-.08em
    T\kern-.1667em\lower.7ex\hbox{E}\kern-.125emX}}
\begin{document}
\title{Output-Feedback Stabilizing Policy Iteration for Convergence Assurance of Unknown Discrete-Time Systems with Unmeasurable States}
\author{Dongdong Li, Jiuxiang Dong
\thanks{Dongdong Li and Jiuxiang Dong are with the College of Information Science and Engineering, and
the State Key Laboratory of Synthetical Automation of Process Industries, Northeastern University, Shenyang 110819, China. Email: lidongdongyq@163.com, dongjiuxiang@ise.neu.edu.cn.
(\emph{Corresponding author: Jiuxiang Dong})}}

\maketitle

\begin{abstract}
This note proposes a data-driven output-feedback stabilizing policy iteration for unknown linear discrete-time systems with unmeasurable states. Existing policy iteration methods for optimal control must start from a stabilizing control policy, which is particularly challenging to obtain for unknown systems, especially when states are unavailable. In such cases, it is more difficult to guarantee stability and convergence performance. To address this problem, an output-feedback stabilizing policy iteration framework is developed to learn closed-loop stabilizing control policies while ensuring convergence performance. Specifically, cumulative scalar parameters are introduced to compress the original system to a stable scale. Then, by integrating modified policy iteration with parameter update rules, the system is gradually amplified/restored  to the original system while preserving stability such that the stabilizing control policy is obtained. The entire process is driven solely by input-output data. Moreover, a stability analysis is provided for output-feedback. The proposed approach is validated by simulations.
\end{abstract}

\section{Introduction}
Reinforcement learning \cite{sutton1998introduction}, or adaptive dynamic programming \cite{lewis2012optimal}, provides effective approach for achieving feedback control with prescribed optimal performance through agent-environment interaction based intelligent decision-making \cite{li2025cooperative}. Current reinforcement learning approaches primarily rely on two iterative frameworks: policy iteration (PI) \cite{1099755,5439950,11087681,7001601,8626769,huang2024specified,11044871} and value iteration \cite{5439950,bian2016value,9852188,huang2024specified}. PI achieves rapid convergence but requires a known initial stabilizing control policy, which may be unsuitable for unstable systems with unknown dynamics. Value iteration, notably, does not require any initial stabilizing policy \cite{bian2016value,11087681}. However, most value iteration approaches require substantial iterations before the value function converges, and their intermediate control policies generally cannot guarantee closed-loop stability throughout the iterative process \cite{11087681}, unlike PI framework. Removing this initial constraint would significantly advance the development and applications of PI. Therefore, this paper aims to develop a model-free output-feedback stabilizing controller design for discrete-time (DT) systems based on the PI framework.

Recently, research on PI has extended beyond the realm of state-feedback control to encompass output-feedback control for the case of unmeasurable states. In practice, the output-feedback framework introduces additional design difficulties when seeking optimal solution through reinforcement learning, including both PI and value iteration. Specifically, in output-feedback scenarios, the initial stabilizing control matrix is formed by multiplying a state-feedback control matrix with a parameterization matrix containing numerous columns, resulting in an output-feedback control matrix with significantly higher dimensions than its state-feedback counterpart (see \cite{7001601,huang2024specified,9769938,jiang2024fully,rizvi2018output}). Moreover, constructing the state-feedback control matrix already presents considerable difficulty, while designing this parameterization matrix further compounds the design complexity. The traditional methods for designing this parameterization matrix heavily depend on prior model knowledge, and the large-scale control matrix may lead to design challenges when approximating initial stabilizing policies \cite{11087681}. To avoid using expanded initial control matrices for initialization, an output-feedback stabilization approach is proposed in \cite{10606096}, whose dimension is reduced to the product of system input and output dimensions, making it considerably smaller than the expanded control matrix. However, this work assumes static output-feedback stabilizability, whose solution may exist after applying additional model constraints. Note that the above PI-based approaches still fundamentally depend on the initial stabilizing control policies.

Recent significant advances have been made in addressing the initial stabilizing constraint of PI. Hybrid iteration methods are proposed in \cite{gao2022resilient,10388382,qasem2023adaptive}, where value iteration is first employed to relax the constraint for an initial stable control policy, followed by policy iteration to achieve rapid convergence. Furthermore, a series of methods have been developed to relax the initial stability condition for PI, such as bias-PI \cite{jiang2022bias} and $\lambda$-PI \cite{9511623}.
These methods relax the initial conditions for PI by using properties of value iteration, but also are affected by their slow convergence, and the stabilizing controller design problem is not completely solved. In \cite{lamperski2020computing}, a discounted PI algorithm is introduced to calculate state-feedback control policies for DT linear systems, however, the update rule for the discount factor is not clear. For continuous-time systems, the homotopy-based PI algorithms are proposed in \cite{chen2022homotopic,10342780,10887315} to compute stabilizing control policies. In \cite{PANG2025112227}, a scaling PI algorithm is proposed for DT linear systems to address the control policy design. Also for DT systems, stabilizing PI (SPI) algorithms are presented in \cite{li2024data,li2025cooperative} that can obtain the initial control policy and achieve optimal control. Although these methods have established mature frameworks for computing stabilizing control policies for continuous-time or DT unknown systems, their applicability remains confined to the full-state feedback case. Due to the aforementioned limitations, they cannot be directly extended to output-feedback control. For instance, extending our proposed state-feedback SPI method \cite{li2024data,li2025cooperative} to output-feedback scenarios would introduce additional challenges, including state reconstruction, iteration step-size design, and stability analysis.
Recently, an output-feedback homotopy-based PI algorithm was proposed in \cite{11087681}, achieving the design of stabilizing control policies for unknown continuous-time linear systems with unmeasurable states. Unlike continuous-time systems where the maximum closed-loop pole/eigenvalue is controlled to the left half-plane of the complex plane, the objective of this paper is to constrain the maximum closed-loop eigenvalue of DT systems within the unit circle. Consequently, the method in \cite{11087681} cannot be directly applied to designing stabilizing control policies for DT systems. Currently, the scarcity of methods for designing stabilizing control policies for DT systems with unmeasurable states significantly hinders the extension and application of output-feedback PI algorithms. Thus, the development of novel algorithms is urgently required to resolve this problem.

Inspired by the above problems and research results, we propose a novel output-feedback SPI algorithm for DT systems with unmeasurable states and unknown model parameters. The main contributions are as follows:

(1). Compared with the existing DT output-feedback PI methods \cite{huang2024specified,9769938,7001601}, the proposed SPI algorithm can obtain an output-feedback stabilizing control policy online. Specifically, we design a mechanism to modify PI by compressing and then restoring the scale of a stabilizing virtual system. This mechanism is regulated by cumulative scalar parameters and, together with the modified policy evaluation-improvement processes, constitutes the output-feedback SPI framework.

(2). Unlike full-state feedback method \cite{li2024data,li2025cooperative}, we construct a extended state vector based on the Kronecker product to achieve system state reconstruction and develop a fully input-output data-driven version for policy evaluation and policy improvement. Moreover, a model-free step-size/parameter selection method and the corresponding stability analysis are established by designing input-output data-driven inequality conditions.

(3). The proposed output-feedback SPI algorithm generates a stabilizing control policy by adjusting scalar parameters, without designing observers or large-scale parameterization matrix. Moreover, the closed-loop stability of the virtual system is guaranteed at each iteration, and the learned output-feedback control policy ensures the convergence performance of the original actual system.

{\it Notations:}
$\otimes$ denotes the Kronecker product. For any matrix $Y\in\mathbb{R}^{a\times b}$, we define $\mathrm{vec}(Y)=[Y_{1}^{\top},Y_{2}^{\top},\ldots,Y_{b}^{\top}]^{\top}\in\mathbb{R}^{ab}$, where $Y_{j}\in\mathbb{R}^{a}$ is $i$th column of matrix $Y$ for $j=1,\ldots,b$. If $Y=Y^{\top}\in\mathbb{R}^{a\times a}$, $\mathrm{vecs}(Y)=[Y_{1,1},2Y_{1,2},\ldots,2Y_{1,a},Y_{2,2},2Y_{2,3}, \ldots,2Y_{a-1,a},Y_{a,a}]^{T}\in\mathbb{R}^{\frac{a(a+1)}{2}}$, where $Y_{i,j}$ is the $i$th row and $j$th column element of matrix $Y$. For vector $v\in\mathbb{R}^{a}$, $\mathrm{vecv}(v)= [v_{1}^{2},v_{1}v_{2},\ldots,v_{1}v_{n},v_{2}^{2},v_{2}v_{3},\ldots,v_{a-1}v_{a},v_{a}^{2}]^{\top}\in\mathbb{R}^{\frac{a(a+1)}{2}}$.

\section{Problem and Preliminaries}

\subsection{Problem Formulation}\label{section:2:1}
A class of DT linear systems with unknown dynamic matrices is considered as follows
\begin{subequations}\label{1}
\begin{align}
x(k+1)&=Ax(k)+Bu(k),\label{1a}\\
y(k)&=Cx(k),\label{1b}
\end{align}
\end{subequations}
where $x(k)\in\mathbb{R}^{n}$, $u(k)\in\mathbb{R}^{m}$ and $y(k)\in\mathbb{R}^{p}$ are the state, input and output, respectively, $A\in\mathbb{R}^{n\times n}$, $B\in\mathbb{R}^{n\times m}$ and $C\in\mathbb{R}^{p\times n}$ are constant matrices with $\mathrm{rank}(C)=\min(p,n)$. The all system matrices are unknown and only the output $y(k)$ in \eqref{1b} is measurable. In other words, only the output can be used for feedback, while the state in \eqref{1a} is not available.
The control objective is to find an output-feedback control policy $u(k)$ such that the state of system \eqref{1} is stabilized, i.e., $\lim_{k\to\infty}x(k)=0$.

Existing output-feedback PI methods, such as \cite{huang2024specified,9769938,7001601}, can achieve optimal control in the case of state unmeasurable, however, they rely on an initial stabilizing control policy. Designing an output-feedback stabilizing control policy for system \eqref{1} with unmeasurable states is difficult. Specifically, the output-feedback stabilizing control gain is composed of the state-feedback stabilizing gain multiplied by a parameterization matrix. As seen from numerous classical output-feedback results \cite{7001601,huang2024specified,9769938,jiang2024fully,rizvi2018output}, the size of this parameterization matrix grows exponentially with the dimensionality of the system state and depends on a priori knowledge of the system model. Designing a state-feedback stabilizing control policy is already challenging, and constructing a large-scale parameterization matrix without utilizing model information is even more difficult. Therefore, this paper will adopt a direct data-driven learning approach to generate the desired output-feedback stabilizing control policy.

If the control objective of this paper is achieved, it can provide a directly usable stabilizing control policy for existing output-feedback PI algorithms, which in turn can solve the linear quadratic regulation problem. Considering the convergence performance of system \eqref{1} and the fitness with the output-feedback PI method, the control objective of this paper can be summarized as the following problem.

{\bf Problem 1.} {\it For the DT system \eqref{1} with unknown dynamics $(A,B,C)$, only the input-output data are used to design an output-feedback control policy $u(k)$ to approximate the behavior of the full-state feedback control law,
\begin{equation} \label{2}
\begin{aligned}
u(k)=-Kx(k),
\end{aligned}
\end{equation}
such that the closed-loop system satisfies
\begin{equation} \label{3}
\begin{aligned}
\rho(A-BK)<1,
\end{aligned}
\end{equation}
where $\rho(\cdot)$ denotes the matrix spectral radius.}

The control law \eqref{2} relies on full-state feedback. However, this paper aims to approximate its behavior using only input-output data $(u(k), y(k))$, without accessing the full-state measurements or the system matrices $(A,B,C)$. The solution to Problem 1 ensures the convergence performance \eqref{3} is met, even when the system states are unmeasurable. To guarantee the solvability of Problem 1, Assumption \ref{A1} is introduced.
\begin{assumption}\label{A1}
The system \eqref{1} is controllable and observable.
\end{assumption}

Based on the above analysis, it follows that addressing Problem 1 or achieving the control objective will face three primary challenges: (i). {\it how to design a basic state-feedback framework for generating stabilizing control policies?} (ii). {\it how to avoid a priori knowledge and reduce initialization complexity, especially to avoid designing state observers or parameterization matrices?} (iii). {\it how to design an input-output data-driven online learning version while ensuring its effectiveness and stability?} These three challenges will be addressed sequentially in subsequent sections, thereby yielding the dynamic output-feedback stabilizing control law that satisfies Problem 1.

\subsection{Preliminaries on State-Feedback SPI}\label{section:2:2}
For system \eqref{1}, there exists a positive constant $\beta$ satisfying
\begin{equation} \label{4}
\begin{aligned}
0<\beta<\tilde{\rho}:=\frac{1}{\rho(A-BK^{0})}
\end{aligned}
\end{equation}
such that $\rho((\tilde{\rho}-\beta)(A-BK^{0}))<1$, where $K^{0}$ is the arbitrarily constant state-feedback control matrix.

A basic SPI framework was proposed in \cite{li2024data} to obtain a state-feedback stabilizing control policy. Nevertheless, this approach necessitates access to the full system state $x(k)$, which confines the algorithm to the state-feedback case.
To better introduce the SPI method, it is first assumed that the system matrices/states $(A,B,C,x(k))$ in \eqref{1} are available. Then, the full-state-feedback SPI algorithm \cite{li2024data} is given by the following Lemma \ref{L1}.
\begin{lemma}\label{L1}
Given arbitrarily bounded policy $K_{i}^0$, matrices $R\succ0$ and $Q_{y}=C^{\top}QC$ with $Q\succ0$, constants $\tilde{\beta}$ and $\alpha^{0}$ satisfying $\tilde{\beta}:=\tilde{\rho}-\beta$ and $\tilde{\rho}>\beta>\alpha^0\geq0$.

({\it Policy Evaluation:}) For $j=0,1,2\ldots$, solve $P^{j}$ by
\begin{equation}\label{5}
\begin{aligned}
&\tilde{A}^{j\top}{P}^{j}\tilde{A}^{j}-{P}^{j}=-Q_{y}-{K}^{j\top}R{K}^{j},
\end{aligned}
\end{equation}
where $\tilde{A}^{j}:=(\tilde{\beta}+\sum_{m=0}^{j}\alpha^{m})(A-B{K}^{j})$.

({\it Policy Improvement:}) Update control policy ${K}^{j+1}$ by
\begin{equation}\label{6}
\begin{aligned}
{K}^{j+1}=&(\tilde{\beta}+\sum_{m=0}^{j}\alpha^{m})^{2}\big(R+(\tilde{\beta}+\sum_{m=0}^{j}\alpha^{m})^{2}B^{\top}{P}^{j}B\big)^{-1}\\
&\times B^{\top}{P}^{j}A,
\end{aligned}
\end{equation}
and update iteration step-size $\alpha^{j+1}$ by
\begin{equation}\label{7}
\begin{aligned}
0<\alpha^{j+1}<\frac{1}{\rho(A-B{K}^{j+1})}-(\tilde{\beta}+\sum_{m=0}^{j}\alpha^{m}).
\end{aligned}
\end{equation}
Then, $\rho(A-BK^{j})<1/(\tilde{\beta}+\sum_{m=0}^{j}\alpha^{m})$ holds for each $j=0,1,2,\ldots$.
\end{lemma}

Lemma \ref{L1} presents a generalized version of the classical PI in \cite{1099755}, which incorporates a ``compression and gradual amplification'' mechanism. The original PI \cite{1099755} is recovered by replacing the cumulative coefficient $\tilde{\beta}+\sum_{m=0}^{j}\alpha^{m}$ as $1$ and omitting \eqref{7}. The core idea is to first scale down the original system dynamics by an initial factor $\tilde{\beta} + \alpha^0$, which ensures the compressed system $(\tilde{\beta} + \alpha^0)(A - BK^0)$ is Schur stable. Then, through iterative updates of the control gain and the step-size $\alpha^{j+1}$, the cumulative coefficient $\tilde{\beta} + \sum_{m=0}^{j} \alpha^{m}$ is monotonically increased. This process gradually expands the system dynamics back toward its original scale, while continuously maintaining closed-loop stability. The procedure guarantees that once the cumulative coefficient reaches $\tilde{\beta}+\sum_{m=0}^{j}\alpha^{m}\geq1$, the resulting control policy stabilizes the original system, achieving the design objective.

It should be noted that the method in Lemma \ref{L1} is model-based and depends on the full-state feedback controller \eqref{2}, as indicated by \eqref{5}-\eqref{7}. However, it can serve as a foundational framework for overcoming challenge (i). Extending this framework to the output-feedback case is challenging, as it requires not only input-output data-driven policy evaluation-improvement but also addressing issues such as state reconstruction, iterative step-size design, and stability analysis. These issues correspond to challenges (ii) and (iii), and will be addressed in the next section, where the output-feedback SPI method is proposed.

\section{Output-Feedback SPI Method}
To overcome the challenges (ii) and (iii), this section develops a data-driven output-feedback SPI algorithm. The design encompasses three main components: state reconstruction from input-output data, reformulation of the SPI, and a model-free step-size selection mechanism, which together ensure closed-loop stability.
\subsection{State Reconstruction by Input-Output Data }

To reconstruct the states, the following DT states $r_{y}(k)\in\mathbb{R}^{np}$ and $r_{u}(k)\in\mathbb{R}^{nm}$ are generated by
\begin{subequations}\label{8}
\begin{align}
r_{u}(k+1)&=(I_{p}\otimes M_{r})r_{u}(k)+u(k)\otimes b,\label{8a}\\
r_{y}(k+1)&=(I_{m}\otimes M_{r})r_{y}(k)+y(k)\otimes b,\label{8b}
\end{align}
\end{subequations}
where $r_{y}(0)=0$, $r_{u}(0)=0$, $b=[0,0,\ldots,0,1]^{\top}$ and $M_{r}=\left[
\begin{matrix}
0&1&0&\ldots&0\\
0&0&1&\ldots&0\\
\vdots&\vdots&\vdots&\ddots&\vdots\\
-d_{n}&-d_{n-1}&-d_{n-2}&\ldots&-d_{1}
\end{matrix}
\right]$ with $d_{i}$, for $i=1,2,\ldots,n$, designed to make $M_{r}$ Schur.
\begin{lemma}\cite{9769938}\label{L2}
Under Assumption \ref{A1}, there exist a full row-rank parameterization matrix $\bar{M}\in\mathbb{R}^{n\times n_{r}}$ such that
\begin{equation}\label{9}
\begin{aligned}
x(k)=\bar{M}\bar{r}(k)+w(k),
\end{aligned}
\end{equation}
where $\bar{r}(k)=[r_{u}^{\top}(k),r_{y}^{\top}(k)]^{\top}\in\mathbb{R}^{n_{r}}$ is defined in \eqref{8} and $w(k)=(A-LC)^{k}x(0)$ with $n_{r}=np+nm$ and $L$ being the observer gain.
\end{lemma}
\begin{remark}\label{R1}
Since the pair $(A, C)$ is observable and $M_{r}$ is Schur, an appropriate matrix ${L}$ can be designed such that the eigenvalues of $A-LC$ are equal to those of $M_r$ \cite{9769938}. Then, the error $w(k)$ in \eqref{9} converges exponentially to zero and the rate of convergence is determined by the eigenvalues of $M_{r}$. It is important to note that gain $L$ is only used for analysis and is never used in the learning and control process. According to \eqref{9} and the convergence of the error $w(k)$, the state $x(k)$ can be re-represented by $\bar{M}\bar{r}(k)$ when $\rho(M_{r})\to 0$ or $k\to\infty$, and then the full-state feedback control law \eqref{2} can be approximated as $u(k)=-\bar{K}\bar{r}(k)$ to solve Problem 1, where $\bar{K}=K\bar{M}$. From \cite{9769938}, designing matrix $\bar{M}$ depends on a priori knowledge of the matrices $(A, B, C)$, and in order to avoid this, we integrate $\bar{M}$ into the gain $\bar{K}$, which is obtained by an input-output data-driven learning in the subsequent sections.
\end{remark}
\begin{remark}
The construction of matrix $M_{r}$ is straightforward and follows a well-defined procedure. It is formed as a companion matrix whose last row consists of the coefficients of any prescribed monic real-coefficient Schur-stable polynomial $\lambda^{n}+d_{1}\lambda^{n-1}+\ldots+d_{n-1}\lambda+d_{n}$. This design offers significant flexibility, as any set of eigenvalues lying strictly within the unit circle can be arbitrarily assigned to $M_{r}$, provided they are selected as real numbers or in complex conjugate pairs. This simple yet powerful structure not only facilitates easy implementation but also ensures the exponential convergence of the state reconstruction error $w(k)$ to $0$.
\end{remark}

\subsection{Output-Feedback SPI Design and Analysis}
It is noteworthy that within the output-feedback framework, the output weighting matrix $Q_{y}=C^{\top}QC$ may only be positive semi-definite (for example, $\mathrm{rank}(C)=p<n$). This can potentially challenge the guaranteed positive definiteness of the solution $P_{j}$ to the Lyapunov equation \eqref{5} when the control gain $K^{j}$ is not yet effective at exciting all system modes. To ensure robust numerical stability and convergence properties for the algorithm, we introduce a positive definite regularization matrix $Q_{c}\succ0$. Consequently, the policy evaluation (Lyapunov equation \eqref{5}) is modified as
\begin{equation}\label{10a}
\begin{aligned}
&\tilde{A}^{j\top}{P}^{j}\tilde{A}^{j}-{P}^{j}=-(Q_{y}+Q_{c})-{K}^{j\top}R{K}^{j}.
\end{aligned}
\end{equation}
The role of $Q_{c}$ is to impose a penalty on the full state, thereby mathematically guaranteeing the positive definiteness of the composite cost function weight. Crucially, this modification preserves the stabilizing properties of the original algorithm stated in Lemma \ref{L1}, as it merely strengthens the cost function without altering the underlying problem structure.

First, rewrite system \eqref{1a} as
\begin{equation} \label{10}
\begin{aligned}
x(k+1)=(\tilde{\beta}+\sum_{m=0}^{j}\alpha^{m})^{-1}\tilde{A}^{j}x(k)+B(K^{j}x(k)+u(k)).
\end{aligned}
\end{equation}
By using \eqref{10a}, \eqref{10} and the definition of $\tilde{A}^{j}$, we have
\begin{equation} \label{11}
\begin{aligned}
&x^{\top}(k+1)P^{j}x(k+1)-x^{\top}(k)P^{j}x(k)\\
=&-(\tilde{\beta}+\sum_{m=0}^{j}\alpha^{m})^{-2}x(k)^{\top}(Q_{y}+{K}^{j\top}R{K}^{j}+Q_{c})x(k)\\
&-[1-(\tilde{\beta}+\sum_{m=0}^{j}\alpha^{m})^{-2}]x^{\top}(k){P}^{j}x(k)\\
&+2x(k)^{\top}A^{\top}{P}^{j}B({K}^{j}x(k)+u(k))+u^{\top}(k)B^{\top}{P}^{j}Bu(k)\\
&-x^{\top}(k){K}^{j\top}B^{\top}{P}^{j}B{K}^{j}x(k).
\end{aligned}
\end{equation}
Substituting the reconstructed state in \eqref{9} into \eqref{11} yields
\begin{equation} \label{12}
\begin{aligned}\nonumber
&r^{\top}(k+1)\bar{P}^{j}r(k+1)-r^{\top}(k)\bar{P}^{j}r(k)\\
\end{aligned}
\end{equation}
\begin{equation} \label{12}
\begin{aligned}
=&-(\tilde{\beta}+\sum_{m=0}^{j}\alpha^{m})^{-2}\bigg(y^{\top}(k)Qy(k)+r^{\top}(k)(\bar{K}^{j\top}R\bar{K}^{j}\\
&+\bar{Q}_{c})r(k)\bigg)-\big(1-(\tilde{\beta}+\sum_{m=0}^{j}\alpha^{m})^{-2}\big)r^{\top}(k)\bar{P}^{j}r(k)\\
&+u^{\top}(k)B^{\top}{P}^{j}Bu(k)-r^{\top}(k)\bar{K}^{j\top}B^{\top}{P}^{j}B\bar{K}^{j}r(k)\\
&+2r^{\top}(k)\bar{M}^{\top}A^{\top}{P}^{j}B(\bar{K}^{j}r(k)+u(k))+\chi^{j}(k),
\end{aligned}
\end{equation}
where
\begin{equation} \label{12a}
\begin{aligned}
\bar{P}^{j}=\bar{M}^{\top}P^{j}\bar{M}, \quad \bar{K}^{j}=K^{j}\bar{M}, \quad \bar{Q}_{c}=\bar{M}^{\top}Q_{c}\bar{M}
\end{aligned}
\end{equation}
and
$\chi^{j}(k)=-2w^{\top}(k+1)P^{j}\bar{M}\bar{r}(k+1)-w^{\top}(k+1)P^{j}w(k+1)-2w^{\top}(k)P^{j}\bar{M}\bar{r}(k)-w^{\top}(k)P^{j}w(k)
-(\tilde{\beta}+\sum_{m=0}^{j}\alpha^{m})^{-2}[2w^{\top}(k){K}^{j\top}R\bar{K}^{j}r(k)+w^{\top}(k){K}^{j\top}R{K}^{j}w(k)
+2w^{\top}(k)Q_{c}\bar{M}r(k)+w^{\top}(k)Q_{c}w(k)]
-(1-(\tilde{\beta}+\sum_{m=0}^{j}\alpha^{m})^{-2})[w^{\top}(k)P^{j}w(k)+2w^{\top}(k)P^{j}\bar{M}\bar{r}(k)]
-2w^{\top}(k){K}^{j\top}B^{\top}{P}^{j}B\bar{K}^{j}r(k)-w^{\top}(k){K}^{j\top}B^{\top}{P}^{j}B{K}^{j}w(k)
+2w^{\top}(k)A^{\top}{P}^{j}B(\bar{K}^{j}r(k)+u(k))+2w^{\top}(k)A^{\top}{P}^{j}B{K}^{j}w(k)+2r^{\top}(k)\bar{M}^{\top}A^{\top}{P}^{j}B{K}^{j}u(k)$.
Based on Lemma \ref{L2}$, \chi^{j}(k)$ is the overall reconstruction error in the learning process, consisting of the error $w(k)$, the constant matrices and the reconstruction state, which converges exponentially to $0$ as $k\to\infty$.

Collect input-output data matrices in time interval $[k_{0},k_{s}]$:
\begin{equation}\label{13a}
\begin{aligned}
\mathcal{C}_{r}&=[\mathrm{vecv}(r(k_{1})-r(k_{0})),\ldots,\mathrm{vecv}(r(k_{s})-r(k_{s-1}))]^{\top},\\
\mathcal{D}_{\bar{K}^{j}r}&=[\mathrm{vecv}(\bar{K}^{j}r(k_{0})),\ldots,\mathrm{vecv}(\bar{K}^{j}r(k_{s-1}))]^{\top},\\
\mathcal{D}_{rr}&=[r(k_{0})\otimes r(k_{0}),\ldots,r(k_{s-1})\otimes r(k_{s-1})]^{\top},\\
\mathcal{D}_{ur}&=[u(k_{0})\otimes r(k_{0}),\ldots,u(k_{s-1})\otimes r(k_{s-1})]^{\top},\\
\mathcal{D}_{yy}&=[y(k_{0})\otimes y(k_{0}),\ldots,y(k_{s-1})\otimes y(k_{s-1})]^{\top},\\
\mathcal{D}_{r}&=[\mathrm{vecv}(r(k_{0})),\ldots,\mathrm{vecv}(r(k_{s-1}))]^{\top},\\
\mathcal{D}_{u}&=[\mathrm{vecv}(u(k_{0})),\ldots,\mathrm{vecv}(u(k_{s-1}))]^{\top},\\
\mathcal{D}_{\chi^{j}}&=[\chi^{j}(k_{0}),\ldots,\chi^{j}(k_{s-1})]^{\top},
\end{aligned}
\end{equation}
By these data matrices, \eqref{12} can be rewritten as
\begin{equation} \label{13}
\begin{aligned}
\psi^{j}[\mathrm{vecs}({\bar{P}}^{j})^{\top}, \mathrm{vec}({Y}_{1}^{j})^{\top}, \mathrm{vecs}({Y}_{2}^{j})^{\top}]^{\top}=\phi^{j}+\mathcal{D}_{\chi^{j}},
\end{aligned}
\end{equation}
where
\begin{equation}
\begin{aligned}\nonumber
Y_{1}^{j}=&\bar{M}^{\top}A^{\top}{P}^{j}B,\quad Y_{2}^{j}=B^{\top}{P}^{j}B,\\
\psi^{j}=[&\mathcal{C}_{r}+\big(1-(\tilde{\beta}+\sum_{m=0}^{j}\alpha^{m})^{-2}\big)\mathcal{D}_{r},\\
&-2\mathcal{D}_{rr}(I\otimes \bar{K}^{j\top})-2\mathcal{D}_{ur},-\mathcal{D}_{u}+\mathcal{D}_{\bar{K}^{j}r}],\\
\phi^{j}=&-(\tilde{\beta}+\sum_{m=0}^{j}\alpha^{m})^{-2}\big(\mathcal{D}_{yy}\mathrm{vec}(Q)\\
&+\mathcal{D}_{rr}\mathrm{vec}(\bar{K}^{j\top}R\bar{K}^{j}+\bar{Q}_{c})\big).
\end{aligned}
\end{equation}

From \eqref{9}, if the zero initial condition is satisfied, i.e., $x(0)=0$, then $w(k)$ and $\mathcal{D}_{\chi^{j}}$ in \eqref{13} are equal to zero. According to \cite[Thm 2]{9769938}, even if the zero initial condition is not satisfied, \eqref{13} also can be approximated as
\begin{equation} \label{14}
\begin{aligned}
\psi^{j}[\mathrm{vecs}(\hat{\bar{P}}^{j})^{\top}, \mathrm{vec}(\hat{Y}_{1}^{j})^{\top}, \mathrm{vecs}(\hat{Y}_{2}^{j})^{\top}]^{\top}=\phi^{j},
\end{aligned}
\end{equation}
where $\hat{\bar{P}}^{j}$, $\hat{Y}_{1}^{j}$ and $\hat{Y}_{2}^{j}$ are approximations of the solutions of \eqref{13} to distinguish from them. If the zero initial condition is satisfied, then \eqref{14} is equivalent to \eqref{13}. From the analysis of \cite[Thm 2]{9769938}, it can be seen that the error between the solutions of \eqref{13} and \eqref{14} can be reduced by decreasing $\mathcal{D}_{\chi^{j}}$. As described in Remark \ref{R1}, by setting a sufficiently small spectral radius $\rho(M_{r})$ or a sufficiently large $k_{0}$, a sufficiently small error $\mathcal{D}_{\chi^{j}}$ can be obtained. Thus, it is feasible to utilize \eqref{14} to approximate \eqref{13}.
The following Assumption is given to ensure that \eqref{14} can be solved uniquely.
\begin{assumption}\label{A2}
The collected input-output data satisfy
\begin{equation} \label{15}
\begin{aligned}
\mathrm{rank}([\mathcal{D}_{r},\mathcal{D}_{ur},\mathcal{D}_{u}])=\frac{1}{2}(n_{r}(n_{r}+1)+m(m+1))+n_{r}m.
\end{aligned}
\end{equation}
\end{assumption}

A similar assumption is common in output-feedback adaptive dynamic programming methods \cite{9769938,huang2024specified,11087681} and is a fundamental Persistent Excitation condition. Then, under Assumption \ref{A2}, for $j=0,1,2,\ldots$, the unique solution to \eqref{14} can be obtained by
\begin{equation} \label{16}
\begin{aligned}
&[\mathrm{vecs}(\hat{\bar{P}}^{j})^{\top}, \mathrm{vec}(\hat{Y}_{1}^{j})^{\top}, \mathrm{vecs}(\hat{Y}_{2}^{j})^{\top}]^{\top}=(\psi^{jT}\psi^{j})^{-1}\psi^{jT}\phi^{j}.
\end{aligned}
\end{equation}

\begin{lemma}\label{L3}
Under Assumptions \ref{A1}-\ref{A2}, the output-feedback equation \eqref{14} over time $[k_{0},k_{s}]$, matrices $\hat{\bar{P}}^{j}$, $\hat{Y}_{1}^{j}$ and $\hat{Y}_{2}^{j}$, can be uniquely solved.
\end{lemma}

{\it Proof.}
Proving the uniqueness of the solution of \eqref{14} is equivalent to proving that the equation $\psi^{j}\Pi_{v}=0$ has a unique zero solution $\Pi_{v}=0$, where $\Pi_{v}=[\eta_{p},\eta_{1},\eta_{2}]^{\top}$ with $\eta_{p}=\mathrm{vecs}(\eta^{v}_{p})^{\top}$, $\eta_{1}=\mathrm{vec}(\eta^{v}_{1})^{\top}$, $\eta_{2}=\mathrm{vecs}(\eta^{v}_{2})^{\top}$, $\eta^{v}_{p}=\eta^{v\top}_{p}$, and $\eta^{v}_{2}=\eta^{v\top}_{2}$. Define $\eta^{v}_{p}=\bar{M}^{\top}L^{v}\bar{M}$, where the full-row rank property of $\bar{M}$ is guaranteed by the controllability of Assumption \ref{A1}. Under the zero initial condition $x(0)=0$, the approximation error $\chi^{j}$ in \eqref{12} is zero. Thus, it follows from \eqref{12} that $\psi^{j}\Pi_{v}=0$ implies
\begin{equation} \label{17}
\begin{aligned}
&[\mathcal{D}_{r},\mathcal{D}_{ur},\mathcal{D}_{u}][\mathrm{vecs}(\kappa_{p})^{\top}, \mathrm{vec}(\kappa_{1})^{\top}, \mathrm{vecs}(\kappa_{2})^{\top}]^{\top}=0,
\end{aligned}
\end{equation}
where $\kappa_{p}=\bar{M}^{\top}[\bar{A}^{j\top}L^{v}\bar{A}^{j}-(\tilde{\beta}+\sum_{m=0}^{j}\alpha^{m})^{-2}L^{v}]\bar{M}+\bar{K}^{j\top}(B^{\top}L^{v}B-\eta^{v}_{2})\bar{K}^{j}
+2(\bar{M}^{\top}A^{\top}L^{v}B-\eta^{v}_{1})\bar{K}^{j}$, $\kappa_{1}=\bar{M}^{\top}A^{\top}L^{v}B-\eta^{v}_{1}$, $\kappa_{2}=B^{\top}L^{v}B-\eta^{v}_{2}$ with $\bar{A}^{j}:=A-BK^{j}$.

Under Assumption \ref{A2}, the matrix $[\mathcal{D}_{r},\mathcal{D}_{ur},\mathcal{D}_{u}]$ is full-column rank. Therefore, \eqref{17} can be solved uniquely as $[\mathrm{vecs}(\kappa_{p})^{\top}, \mathrm{vec}(\kappa_{1})^{\top}, \mathrm{vecs}(\kappa_{2})^{\top}]^{\top}=0$. Since $\bar{M}$ is full-row rank matrix, one has that $\bar{A}^{j\top}L^{v}\bar{A}^{j}-(\tilde{\beta}+\sum_{m=0}^{j}\alpha^{m})^{-2}L^{v}=0$, where $(\tilde{\beta}+\sum_{m=0}^{j}\alpha^{m})^{2}\bar{A}^{j}=\tilde{A}^{j}$ is Schur. Thus, $L^{v}$ must be zero, then $\eta^{v}_{p}$ also must be zero. Moreover, there exist $\eta^{v}_{1}=\bar{M}^{\top}A^{\top}L^{v}B=0$ and $\eta^{v}_{2}=B^{\top}L^{v}B=0$. Finally, we can obtain that $\Pi_{v}=0$. The proof is completed.$\Box$

Based on solution \eqref{16}, the output-feedback control gain can be updated by
\begin{equation} \label{18}
\begin{aligned}
\hat{\bar{K}}^{j+1}=(\tilde{\beta}+\sum_{m=0}^{j}\alpha^{m})^{2}\big(R+(\tilde{\beta}+\sum_{m=0}^{j}\alpha^{m})^{2}\hat{Y}_{2}^{j}\big)^{-1}\hat{Y}_{1}^{j\top}.
\end{aligned}
\end{equation}
\begin{lemma}\label{L4}
Under Assumptions \ref{A1}-\ref{A2} and Lemma \ref{L3}, the gain ${K}^{j+1}\bar{M}$ in \eqref{12a} can be approximated by matrix $\hat{\bar{K}}^{j+1}$ learned from \eqref{18}.
\end{lemma}

{\it Proof.}
From the analyses in Lemmas \ref{L2}-\ref{L3} and Remark 1, $\hat{Y}_{1}^{j}$ and $\hat{Y}_{2}^{j}$ in \eqref{14} can be made to converge to ${Y}_{1}^{j}$ and ${Y}_{2}^{j}$ in \eqref{13}, respectively, if the time $k_{0}$ is sufficiently large or the spectral radius $\rho(M_{r})$ is sufficiently small. This, together with the uniqueness in Lemma \ref{L3}, leads that the learned matrix $\hat{\bar{K}}^{j+1}$ in \eqref{18} uniquely converges to the gain ${K}^{j+1}\bar{M}$ in \eqref{12a}.
Therefore, gain ${K}^{j+1}\bar{M}$ can be approximated by $\hat{\bar{K}}^{j+1}$.$\Box$

\begin{remark}
In this paper, the initial time $k_0$ is determined by the decay rate of the state reconstruction error $w(k)$, which is governed by the spectral radius $\rho(M_r)$. A practical lower bound is given by $k_0 \geq \ln \epsilon_w / \ln \rho(M_r)$, where $\epsilon_w$ is the desired error tolerance. Although the unavailability of system matrices prevents an exact explicit relation, this criterion provides a theoretically grounded guideline for choosing $k_0$ in practice. Based on the appropriate $k_0$, gain $\hat{\bar{K}}^{j+1}$ learned from \eqref{18} can precisely approximate ${K}^{j+1}\bar{M}$.
\end{remark}
\subsection{Iteration Step-Size Design and Stability Analysis}
Based on the input-output data-driven policy evaluation \eqref{16} and policy improvement \eqref{18}, the remaining crucial issue for the output-feedback SPI is to ensure stability during the iterative learning process. This subsection addresses the fundamental requirement by developing a novel iteration step-size design for $\alpha^{j+1}$  that guarantees the stability of the SPI.

In Lemma 1, the selection scheme \eqref{7} for the iteration step-size $\alpha^{j+1}$ is model-based. To relax this restriction, we choose $\alpha^{j+1}>0$ by solving
\begin{equation} \label{19}
\begin{aligned}
&\big((1+\frac{\alpha^{j+1}}{\tilde{\beta}+\sum_{m=0}^{j}\alpha^{m}})^{2}-1\big)\Pi\leq\Xi,
\end{aligned}
\end{equation}
where
\begin{equation} \label{19a}
\begin{aligned}
\Pi:=&r^{\top}(k)\hat{\bar{P}}^{j}r(k)-r^{\top}(k)(\hat{\bar{K}}^{j+1\top}R\hat{\bar{K}}^{j+1}+\bar{Q}_{c})r(k)\\
&-y^{\top}(k)Qy(k)\\
\Xi:=&(1-\delta)r^{\top}(k)(\hat{\bar{K}}^{j+1\top}R\hat{\bar{K}}^{j+1}+\bar{Q}_{c})r(k)\\
&+(1-\delta)y^{\top}(k)Qy(k),
\end{aligned}
\end{equation}
and $\tilde{\beta}+\sum_{m=0}^{j}\alpha^{m}$, $\hat{\bar{P}}^{j}$, and $\hat{\bar{K}}^{j+1}$ have been solved previously, $\delta$ is a constant coefficient satisfying $0<\delta<1$. Next, we will analyze the solvability of inequality \eqref{19} and the stability.

\begin{lemma}\label{L5}
Given the control gain $\hat{\bar{K}}^{j+1}$ by \eqref{18} and the iteration step-size scheme \eqref{19} for $j=0,1,2,\ldots$, we have that: (i). inequality \eqref{19} has at least one positive scalar solution $\alpha^{j+1}$ ({\it Solvability}); (ii). if update $\alpha^{j+1}$ by \eqref{19},
the closed-loop system $x(k+1)=(\tilde{\beta}+\sum_{m=0}^{j+1}\alpha^{m})(Ax(k)-B\hat{\bar{K}}^{j+1}r(k))$ can approximate the full-state
feedback system $x(k+1)=(\tilde{\beta}+\sum_{m=0}^{j+1}\alpha^{m})(A-BK^{j+1})x(k)$ with
guaranteeing asymptotic convergence of the system state ({\it Stability}).
\end{lemma}

{\it Proof.}
Proof of (i). Based on Lemmas \ref{L3}-\ref{L4}, since ${K}^{j+1}\bar{M}$ is approximated by $\hat{\bar{K}}^{j+1}$, then the controller $-\hat{\bar{K}}^{j+1}r(k)$ approximates $-{K}^{j+1}x(k)$. It follows from \cite{1099755} that given the Schur matrix $\tilde{A}^{j}$, the system $(\tilde{\beta}+\sum_{m=0}^{j}\alpha^{m})(A-B{K}^{j+1})$ can be made Schur if the control gain ${K}^{j+1}$ is updated by \eqref{6}. Thus, there must exist the unique positive solution $\tilde{P}^{j}$ to
\begin{equation} \label{20}
\begin{aligned}
&(\tilde{\beta}+\sum_{m=0}^{j}\alpha^{m})^{2}(A-B{K}^{j+1})^{\top}\tilde{P}^{j}(A-B{K}^{j+1})-\tilde{P}^{j}\\
=&-Q_{y}-Q_{c}-{K}^{j+1\top}R{K}^{j+1},
\end{aligned}
\end{equation}
where matrix $\tilde{P}^{j}$ satisfies $0\prec\tilde{P}^{j}\leq {P}^{j}$.
Clearly, it follows from \eqref{20} that $\tilde{P}^{j}-Q_{y}-Q_{c}-{K}^{j+1\top}R{K}^{j+1}\succ0$. By $0\prec\tilde{P}^{j}\leq {P}^{j}$, one has ${P}^{j}-Q_{y}-Q_{c}-{K}^{j+1\top}R{K}^{j+1}\succ0$. This is equivalent to $x^{\top}(k){P}^{j}x(k)-y^{\top}(k)Qy(k)-x^{\top}(k)({K}^{j+1\top}R{K}^{j+1}+Q_{c})x(k)>0$.  Similar to Lemma \ref{L4}, $\bar{M}^{\top}{P}^{j}\bar{M}$ can be approximated by $\hat{\bar{P}}^{j}$. Thus, if the $w(k)$ is sufficiently small, we can obtain
\begin{equation} \label{21}
\begin{aligned}
x^{\top}(k){P}^{j}x(k)=r^{\top}(k)\hat{\bar{P}}^{j}r(k), {K}^{j+1}x(k)=\hat{\bar{K}}^{j+1}r(k).
\end{aligned}
\end{equation}
This implies $r^{\top}(k)\hat{\bar{P}}^{j}r(k)-r^{\top}(k)(\hat{\bar{K}}^{j+1\top}R\hat{\bar{K}}^{j+1}+\bar{Q}_{c})r(k)-y^{\top}(k)Qy(k)>0$. Since $0<\delta<1$ and $\alpha^{j+1}>0$, inequality \eqref{19} has at least one positive scalar solution $\alpha^{j+1}$.

Proof of (ii).
Consider the Lyapunov function candidate $V(k)=x^{\top}(k)\tilde{P}^{j}x(k)$ with respect to system $x(k+1)=\tilde{A}^{j+1}x(k)$, where $\tilde{A}^{j+1}=(\tilde{\beta}+\sum_{m=0}^{j+1}\alpha^{m})(A-B{K}^{j+1})$. Then, one has
\begin{equation} \label{22}
\begin{aligned}
&V(k+1)-V(k)\\
=&x^{\top}(k)\tilde{A}^{j+1\top}\tilde{P}^{j}\tilde{A}^{j+1}x(k)-x^{\top}(k)\tilde{P}^{j}x(k)\\
\overset{(a)}{=}&\big((\frac{\tilde{\beta}+\sum_{m=0}^{j}\alpha^{m}}{\tilde{\beta}+\sum_{m=0}^{j+1}\alpha^{m}})^{2}-1\big)
x^{\top}(k)\big(\tilde{P}^{j}-{K}^{j+1\top}R{K}^{j+1}-Q_{y}\\
&-Q_{c}\big)x(k)-x^{\top}(k)\big({K}^{j+1\top}R{K}^{j+1}+Q_{y}+Q_{c}\big)x(k),
\end{aligned}
\end{equation}
where $(a)$ is obtained from \eqref{20}. By using $0\prec\tilde{P}^{j}\leq {P}^{j}$ and \eqref{21}, we have
\begin{equation} \label{22}
\begin{aligned}
&V(k+1)-V(k)\\
\leq&\big((1+\frac{\alpha^{j+1}}{\tilde{\beta}+\sum_{m=0}^{j+1}\alpha^{m}})^{2}-1\big)\big(r^{\top}(k)\hat{\bar{P}}^{j}r(k)-y^{\top}(k)Qy(k)\\
&-r^{\top}(k)(\hat{\bar{K}}^{j+1\top}R\hat{\bar{K}}^{j+1}+\bar{Q}_{c})r(k)\big)-y^{\top}(k)Qy(k)\\
&-r^{\top}(k)(\hat{\bar{K}}^{j+1\top}R\hat{\bar{K}}^{j+1}+\bar{Q}_{c})r(k).
\end{aligned}
\end{equation}
substituting \eqref{19} and \eqref{21} into \eqref{22} yields
\begin{equation} \label{23}
\begin{aligned}
&V(k+1)-V(k)\\
\leq&-\delta y^{\top}(k)Qy(k)-\delta x^{\top}(k)(K^{j+1\top}RK^{j+1}+Q_{c})x(k).
\end{aligned}
\end{equation}
Based on \eqref{23} and the observability of $(A,C)$, it can be concluded that the state $x$ asymptotically converges. In other words, if update $\alpha^{j+1}$ by \eqref{19}, the output-feedback controller $-\hat{\bar{K}}^{j+1}r(k)$ can approximate the full-state-feedback controller $-K^{j+1}x(k)$ to make the system $x(k+1)=\tilde{A}^{j+1}x(k)$ asymptotically stable. The proof is completed. $\Box$

\begin{algorithm}[!h]
	\caption{Output-Feedback SPI Algorithm}\label{alg1}
	\begin{algorithmic}[1]
	\STATE {\bf Initialize:} Set arbitrary control policy $\hat{\bar{K}}^{0}$, a monotonically decreasing sequence $\{\tilde{\beta}^{z}\}$ between $0$ and $1$, $\alpha^{0}=0$, and $\bar{Q}_{c}=0$. $j\leftarrow0$, $z\leftarrow0$, $\tilde{\beta}\leftarrow\tilde{\beta}^{0}$.
\STATE  Collect data by \eqref{13a} over the time interval $[k_{0}, k_{s}]$ until Assumption \ref{A2} is satisfied.
\STATE Compute $\hat{\bar{P}}^{0}$ by \eqref{14} with $\tilde{\beta}$ and $\alpha^0$.
\IF{$r^{\top}(k)\hat{\bar{P}}^{0}r(k)\leq0$}
    \STATE $z\leftarrow z+1$
    \STATE $\tilde{\beta}\leftarrow\tilde{\beta}^{z}$ and go to Step 3.
\ELSE
    \STATE Set $\bar{Q}_{c}\gets\hat{\bar{P}}^{0}$.
    \WHILE{$\tilde{\beta}+\sum_{m=0}^{j}\alpha^{m}\geq1$ is not satisfied}
        \STATE Compute $\hat{\bar{K}}^{j+1}$ from \eqref{18}.
        \STATE Select $\alpha^{j+1}$ by \eqref{19}.
        \STATE $j\gets j+1$
    \ENDWHILE
\ENDIF
\RETURN the output-feedback control gain $\bar{K}=\hat{\bar{K}}^{j}$.
	\end{algorithmic}
\end{algorithm}
\begin{theorem}\label{the1}
Consider system \eqref{1} under Assumptions \ref{A1} and \ref{A2}. The output-feedback control law $u(k)=-\bar{K}r(k)$ solves Problem 1 with the convergence performance assurance \eqref{3}, where the gain $\bar{K}$ learned from Algorithm \ref{alg1}.
\end{theorem}

{\it Proof.}
From Lemma \ref{L4}, the implemented controller can be expressed as
$u(k) = -K^{j+1} x(k) + K^{j+1} w(k)$,
where $w(k)$ is the state reconstruction error that converges exponentially to zero. Therefore, for a sufficiently large $k_0$, the controller $u(k) = -\hat{\bar{K}}^{j+1}r(k)$ approximates the full-state feedback law
$u(k) = -K^{j+1} x(k)$.
Under the control law $u(k) = -K^{j+1} x(k)$, the closed-loop system dynamics are given by
$x(k+1) = (A - B K^{j+1}) x(k)$.
From Lemma \ref{L1} and $\tilde{\beta}+\sum_{m=0}^{j+1}\alpha^{m}\geq1$, the spectral radius satisfies $\rho(A - B K^{j+1})<1$, which implies that the closed-loop system is Schur stable. Convergence performance \eqref{3} is satisfied.
$\Box$

\begin{remark}
Note that in Algorithm \ref{alg1}, we introduce a check term $r^{\top}(k) \hat{\bar{P}}^0 r(k)$ to validate and adjust the value of $\tilde{\beta}$. Since $\hat{\bar{P}}^0$ is the solution to the Lyapunov equation, if $r^{\top}(k) \hat{\bar{P}}^0 r(k)> 0$, it indicates that the compressed system $(\tilde{\beta} + \alpha^0)(A-BK)$, with $\alpha^0 = 0$, is initially stable. Conversely, if $r^{\top}(k) \hat{\bar{P}}^0 r(k) \leq 0$, $\tilde{\beta}$ is reduced to further compress the original system. This verification and adjustment mechanism operates without relying on any prior knowledge. Furthermore, assigning the learned positive definite matrix \(\hat{\bar{P}}^{0}\) to \(\bar{Q}_{c}\) satisfies the requirement for \(\bar{Q}_{c} \succ 0\) without necessitating any additional design effort.
\end{remark}
\begin{remark}
The model-free selection scheme \eqref{19} is contained within the model-based scheme \eqref{7}. Theorem \ref{the1} implies that selecting $\alpha^{j+1}$ via \eqref{19} ensures the Schur stability of $\tilde{A}^{j+1}$, i.e., $\rho((\tilde{\beta} + \sum_{m=0}^{j+1} \alpha^{m})(A - BK^{j+1})) < 1$, which is equivalent to $\rho(A - BK^{j+1}) < 1 / (\tilde{\beta} + \sum_{m=0}^{j+1} \alpha^{m})$. Therefore, the $\alpha^{j+1}$ chosen by \eqref{19} satisfies \eqref{7}. Moreover, since $A - B\hat{\bar{K}}^{j}\bar{M}^{+}$ is equivalent to $A - BK^{j}$, where $\bar{M}^{+} = \bar{M}^{\top}(\bar{M}\bar{M}^{\top})^{-1}$ represents the pseudo-inverse of $\bar{M}$, the effectiveness of the algorithm can be verified in simulations by examining the relationship between $1/(\tilde{\beta} + \sum_{m=0}^{j} \alpha^{m})$ and $\rho(A - B\hat{\bar{K}}^{j}\bar{M}^{+})$. Note that the full row-rank property of $\bar{M}$ ensures that $\bar{M}^{+}$ is always exists, but these matrices are only used for simulation verification and is never employed during the learning and control processes.
\end{remark}
\begin{remark}\label{R6}
The coefficient $\delta$ is important for tuning the convergence speed of Algorithm \ref{alg1}. As seen in the selection scheme \eqref{19}, a larger $(1-\delta)$ allows a larger $\alpha^{j+1}$, helping the cumulative coefficient $\tilde{\beta}+\sum_{m=0}^{j}\alpha^{m}$ reach the termination threshold faster. Thus, by adjusting $\delta$, the user can control the speed at which the output-feedback stabilizing control gain is generated.
\end{remark}
\begin{remark}\label{R7}
Existing output-feedback PI algorithms \cite{huang2024specified,9769938,7001601} usually assume that an initial stabilizing control matrix $\hat{\bar{K}}^0$ is known with dimensions $m\times n_{r}$, where $n_{r}=np+nm$. Even when a state-feedback stabilizing control gain $K^0$ in \cite{li2024data,li2025cooperative} has been determined, it is necessary to initialize an additional parameterization matrix $\bar{M}$ with dimensions $n \times n_r$, which depends on prior knowledge of the model. In Algorithm \ref{alg1}, it is not required to carefully design a stabilizing initial control matrix $\hat{\bar{K}}^0\in\mathbb{R}^{m\times n_{r}}$, nor is it required to design a large-scale matrix $\bar{M}\in\mathbb{R}^{n\times n_{r}}$. Instead, only a scalar parameter $\alpha^{j+1}$ with dimension $1$ needs to be adjusted. This overcomes challenge (ii) described in Section \ref{section:2:1}, thereby reducing complexity. In summary, Algorithm \ref{alg1} completely overcomes challenge (iii) and solves Problem 1.
\end{remark}

\section{Simulation Example}
In this section, we utilize a discretized power system given in \cite{PANG2025112227} to validate the proposed output-feedback SPI algorithm, with the closed-loop dynamics described by
\begin{equation} \label{24}
\begin{aligned}
x(k+1) &=
\begin{bmatrix}
0.8825 & 0.0014 & 0.0470 \\
0.0894 & 0.9049 & 0.0023 \\
0.0028 & 0.0571 & 0.9995
\end{bmatrix} x(k) +
\begin{bmatrix}
0.0001 \\
0.1190 \\
0.0036
\end{bmatrix} u,\\
y(k) &= \begin{bmatrix}
1 & 0 &  0
\end{bmatrix} x(k),
\end{aligned}
\end{equation}
where the state $x = [x_1, x_2, x_3]^\top \in \mathbb{R}^3$.
The eigenvalues of the open-loop matrix $A$ are $\lambda_{1,2} = 0.8847 \pm 0.0405i$ and $\lambda_{3} = 1.0176$,
yielding a spectral radius $\rho(A)=1.0176>1$. This implies that the open-loop system is unstable.
\begin{figure}[htbp]
    \centering
	  \subfloat[]{
       \includegraphics[width=0.47\linewidth]{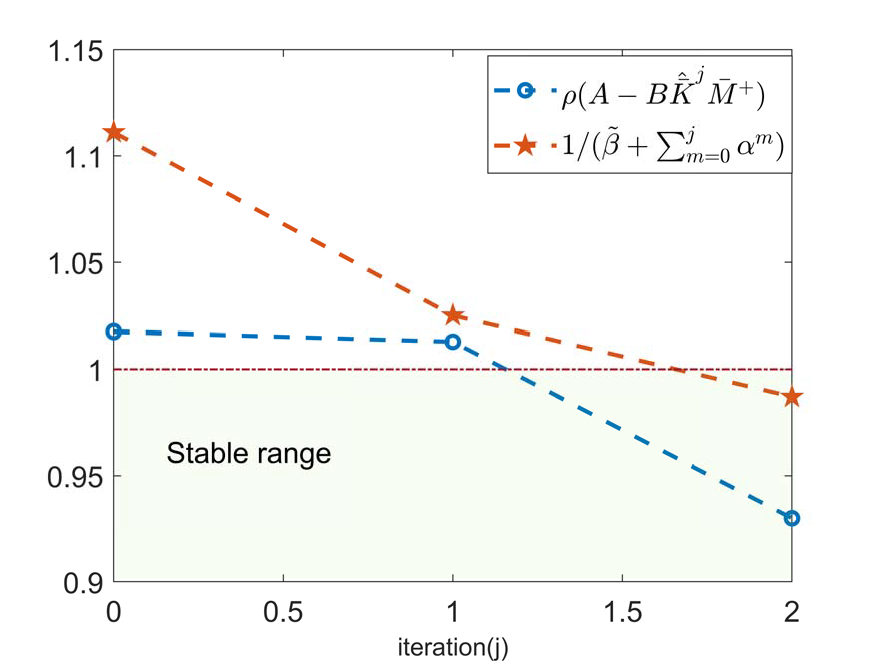}}
       \hfill
	  \subfloat[]{
        \includegraphics[width=0.47\linewidth]{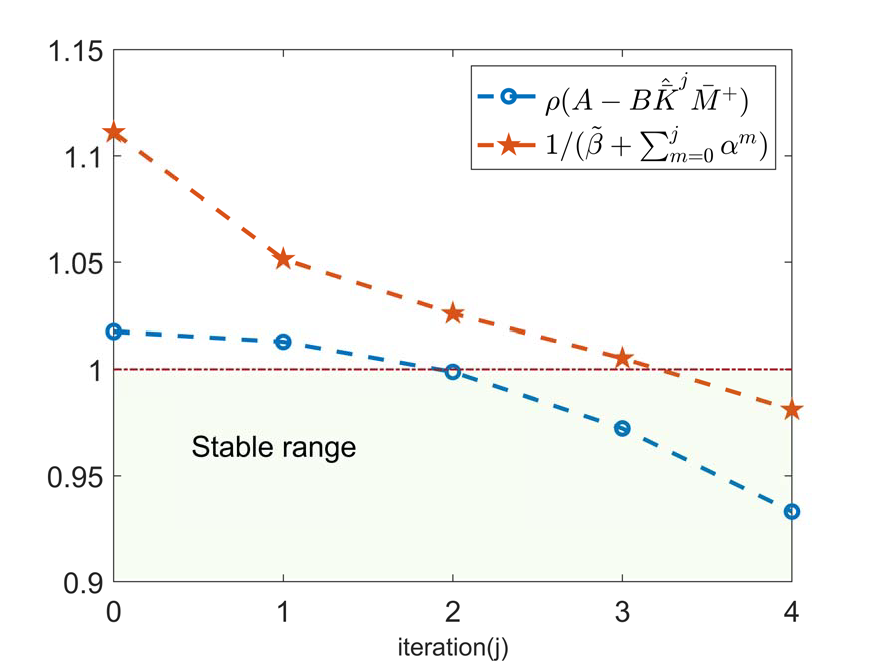}}
        \hfill
	  \subfloat[]{
        \includegraphics[width=0.47\linewidth]{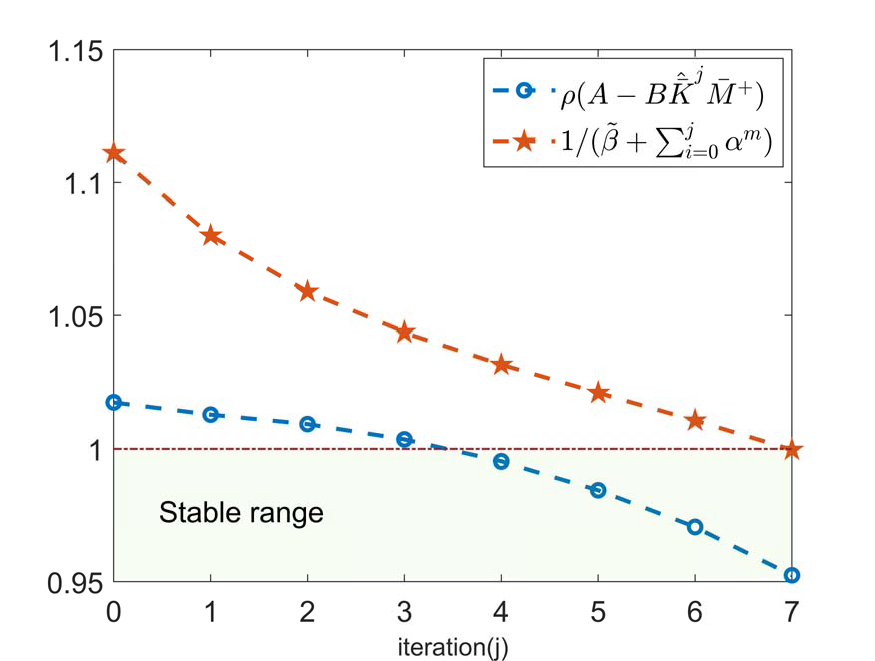}}
        \hfill
	  \subfloat[]{
        \includegraphics[width=0.47\linewidth]{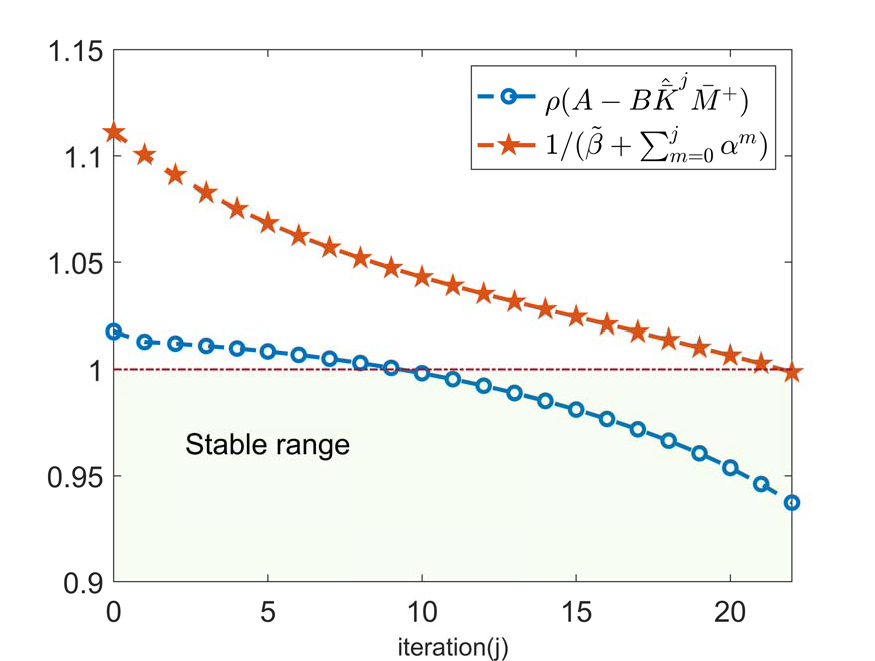}}
	  \caption{Closed-loop spectral radius $\rho(A - B\hat{\bar{K}}^{j}\bar{M}^{+})$ and its upper bound $1/(\tilde{\beta} + \sum_{m=0}^{j} \alpha^{m})$, where $\bar{M}^{+} = \bar{M}^{\top}(\bar{M}\bar{M}^{\top})^{-1}$: (a). $\delta=0.1$; (b). $\delta=0.4$; (c). $\delta=0.7$; (d). $\delta=0.9$.}\label{fig1}
	  \label{fig2}
\end{figure}
\begin{figure}[htbp]
      \centering
      \includegraphics[width=8cm,height=6cm]{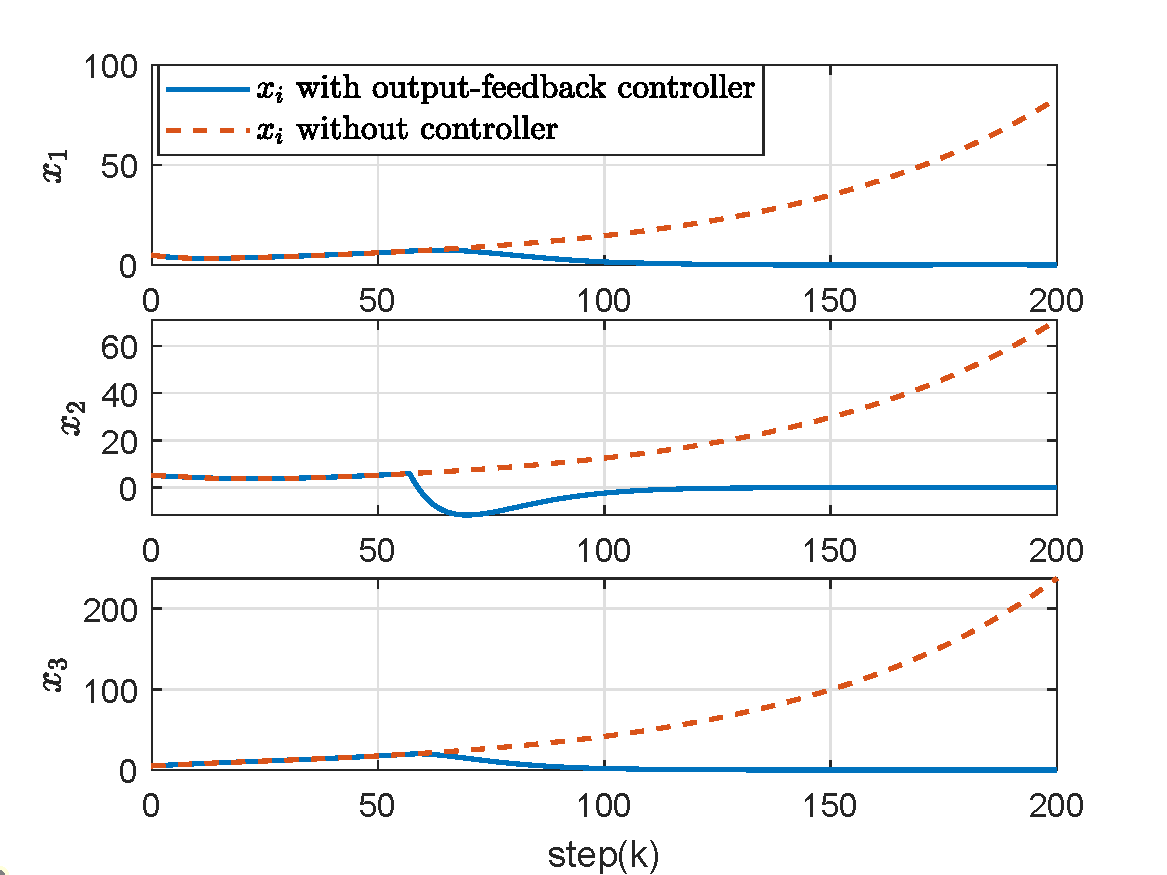}
      \caption{The system state trajectories with learned output-feedback controller and without controller.}\label{fig2}
\end{figure}

The system is both controllable and observable, satisfying Assumption \ref{A1}. For output-feedback implementation, we construct the dynamic states $r_u(k)$ and $r_y(k)$ using the Kronecker-based approach in \eqref{8a}-\eqref{8b}, with $M_r$ designed to be Schur with eigenvalues at $[-0.1, -0.2, -0.3]$. To initialize the algorithm, we set the initial control gain as $\hat{\bar{K}}^0 = 0$, the initial step-size $\alpha^0 = 0$, the monotonically decreasing sequence $\{\tilde{\beta}^{z}\}=\{0.9,0.89,0.88,\ldots,0.01\}$, and the coefficient $\delta=0.7$, and select the weighting matrices as $Q = I_3$ and $R = 1$. The initial state is set to $x(0) = [5, 5, 5]^\top$. Prior to data collection, we inject persistent excitation signals comprising a combination of sinusoidal waves with different frequencies into the system to satisfy the rank condition in Assumption \ref{A2}. By using Algorithm \ref{alg1}, we obtain the stabilizing output-feedback control gain as
 \begin{equation} \label{25}
\begin{aligned}
\hat{\bar{K}}^{7}=[0.0237,\quad	0.0568,\quad	0.0518,\quad	214.7320,&\\
	-487.3737,\quad	277.3016&],
\end{aligned}
\end{equation}
with the closed-loop spectral radius $\rho(A - B\hat{\bar{K}}^{j}\bar{M}^{+})=0.9525$, where $\hat{\bar{K}}^{7}\bar{M}^{+}=[0.4006,0.4080,0.8978]$ is obtained by learning.

To verify the impact of different $\delta$ values on the learning speed of the algorithm, we present the results of the closed-loop spectral radius and its upper bound for $\delta=0.1,0.4,0.7,0.9$ as shown in Figs. \ref{fig1}(a)-\ref{fig1}(d). The actual spectral radius $\rho(A - B\hat{K}^j\bar{M}^+)$ remains consistently below the upper bound $1/(\tilde{\beta} + \sum_{m=0}^{j} \alpha^m)$ across all iterations, confirming the convergence guarantee established in Theorem \ref{the1}. Furthermore, these results validate the discussion in Remark \ref{R6} that a smaller $\delta$ leads to the output-feedback stabilizing control gain being obtained more rapidly.

Fig. \ref{fig2} compares the state trajectories of the uncontrolled system and the system under the learned output-feedback controller $u(k)=-\hat{\bar{K}}^{7}r(k)$. Without control, all three states diverge rapidly due to the open-loop instability. In contrast, when the data-driven output-feedback controller is applied, all states converge to zero, despite the controller having no access to true state measurements. This result validates the practical effectiveness of the proposed algorithm in stabilizing unknown systems using only input-output data.

\section{Conclusion}
This note presents an output-feedback SPI algorithm for discrete-time linear systems with completely unknown dynamics and unmeasurable states. The proposed approach introduces cumulative scalar coefficients to adjust the scale of system dynamics and employs Kronecker product-based dynamic state reconstruction. A complete data-driven framework is established, incorporating policy evaluation, policy improvement, and model-free adaptive step-size selection while maintaining closed-loop stability during the iterations. Theoretical analysis and simulations verify that the proposed algorithm ensures closed-loop stability while learning control policies directly from input-output data. This study provides a solution for generating initial stabilizing policies in output-feedback PI methods, extending their applicability to scenarios with unmeasurable states.


\begin{thebibliography}{10}

\bibitem{sutton1998introduction}
R.~S. Sutton, A.~G. Barto, {\em et~al.}, {\em Introduction to reinforcement
  learning}, vol.~135.
\newblock MIT press Cambridge, 1998.

\bibitem{lewis2012optimal}
F.~L. Lewis, D.~Vrabie, and V.~L. Syrmos, {\em Optimal control}.
\newblock John Wiley \& Sons, 2012.

\bibitem{li2025cooperative}
D.~Li and J.~Dong, ``Cooperative optimal output tracking for discrete-time
  multiagent systems: Stabilizing policy iteration frameworks,'' {\em IEEE
  Trans. Autom. Control}, pp.~1--8, 2025.


\bibitem{1099755}
G.~Hewer, ``An iterative technique for the computation of the steady state
  gains for the discrete optimal regulator,'' {\em IEEE Trans. Autom. Control},
  vol.~16, no.~4, pp.~382--384, 1971.

\bibitem{5439950}
F.~L. Lewis and K.~G. Vamvoudakis, ``Reinforcement learning for partially
  observable dynamic processes: Adaptive dynamic programming using measured
  output data,'' {\em IEEE Trans. Syst. Man Cybern. Part B-Cybern.}, vol.~41,
  no.~1, pp.~14--25, 2011.

\bibitem{11087681}
Y.~Chen, C.~Chen, J.~Kang, F.~L. Lewis, and S.~Xie, ``Output-feedback
  homotopy-based policy iteration for performance assurance of unknown linear
  continuous-time systems,'' {\em IEEE Trans. Autom. Control}, pp.~1--8, 2025.

\bibitem{7001601}
B.~Kiumarsi, F.~L. Lewis, M.-B. Naghibi-Sistani, and A.~Karimpour, ``Optimal
  tracking control of unknown discrete-time linear systems using input-output
  measured data,'' {\em IEEE T. Cybern.}, vol.~45, no.~12, pp.~2770--2779,
  2015.

\bibitem{8626769}
Y.~Jiang, B.~Kiumarsi, J.~Fan, T.~Chai, J.~Li, and F.~L. Lewis, ``Optimal
  output regulation of linear discrete-time systems with unknown dynamics using
  reinforcement learning,'' {\em IEEE T. Cybern.}, vol.~50, no.~7,
  pp.~3147--3156, 2020.

\bibitem{huang2024specified}
C.~Huang, C.~Chen, K.~Xie, F.~L. Lewis, and S.~Xie, ``Specified convergence
  rate guaranteed output tracking of discrete-time systems via reinforcement
  learning,'' {\em Automatica}, vol.~161, p.~111490, 2024.

\bibitem{11044871}
H.~Shen, Y.~Wang, H.~Yan, and S.~Xu, ``Data-driven single-loop policy iteration
  control of uncertain singularly perturbed systems,'' {\em IEEE Trans. Autom.
  Control}, pp.~1--7, 2025.

\bibitem{bian2016value}
T.~Bian and Z.-P. Jiang, ``Value iteration and adaptive dynamic programming for
  data-driven adaptive optimal control design,'' {\em Automatica}, vol.~71,
  pp.~348--360, 2016.

\bibitem{9852188}
C.~Li, J.~Ding, F.~L. Lewis, and T.~Chai, ``Model-free \emph{Q}-learning for
  the tracking problem of linear discrete-time systems,'' {\em IEEE Trans.
  Neural Netw. Learn. Syst.}, vol.~35, no.~3, pp.~3191--3201, 2024.

\bibitem{9769938}
C.~Chen, L.~Xie, Y.~Jiang, K.~Xie, and S.~Xie, ``Robust output regulation and
  reinforcement learning-based output tracking design for unknown linear
  discrete-time systems,'' {\em IEEE Trans. Autom. Control}, vol.~68, no.~4,
  pp.~2391--2398, 2023.

\bibitem{jiang2024fully}
Y.~Jiang, L.~Liu, and G.~Feng, ``Fully distributed adaptive control for output
  consensus of uncertain discrete-time linear multi-agent systems,'' {\em
  Automatica}, vol.~162, p.~111531, 2024.

\bibitem{rizvi2018output}
S.~A.~A. Rizvi and Z.~Lin, ``Output feedback \emph{Q}-learning for
  discrete-time linear zero-sum games with application to the \emph{H}-infinity
  control,'' {\em Automatica}, vol.~95, pp.~213--221, 2018.

\bibitem{10606096}
C.~Zhang, C.~Chen, F.~L. Lewis, and S.~Xie, ``Policy iteration-based learning
  design for linear continuous-time systems under initial stabilizing opfb
  policy,'' {\em IEEE T. Cybern.}, vol.~54, no.~11, pp.~6707--6718, 2024.

\bibitem{gao2022resilient}
W.~Gao, C.~Deng, Y.~Jiang, and Z.-P. Jiang, ``Resilient reinforcement learning
  and robust output regulation under denial-of-service attacks,'' {\em
  Automatica}, vol.~142, p.~110366, 2022.

\bibitem{10388382}
H.~Shen, C.~Peng, H.~Yan, and S.~Xu, ``Data-driven near optimization for fast
  sampling singularly perturbed systems,'' {\em IEEE Trans. Autom. Control},
  vol.~69, no.~7, pp.~4689--4694, 2024.

\bibitem{qasem2023adaptive}
O.~Qasem, W.~Gao, and H.~Gutierrez, ``Adaptive optimal control for
  discrete-time linear systems via hybrid iteration,'' in {\em 2023 IEEE 12th
  Data Driven Control Learn. Syst. Conf. (DDCLS)}, pp.~1141--1146, IEEE, 2023.

\bibitem{jiang2022bias}
H.~Jiang and B.~Zhou, ``Bias-policy iteration based adaptive dynamic
  programming for unknown continuous-time linear systems,'' {\em Automatica},
  vol.~136, p.~110058, 2022.

\bibitem{9511623}
Y.~Yang, B.~Kiumarsi, H.~Modares, and C.~Xu, ``Model-free $\lambda$-policy
  iteration for discrete-time linear quadratic regulation,'' {\em IEEE Trans.
  Neural Netw. Learn. Syst.}, vol.~34, no.~2, pp.~635--649, 2023.

\bibitem{lamperski2020computing}
A.~Lamperski, ``Computing stabilizing linear controllers via policy
  iteration,'' in {\em 2020 IEEE 59th Conf. Decis. Control (CDC)},
  pp.~1902--1907, IEEE, 2020.

\bibitem{chen2022homotopic}
C.~Chen, F.~L. Lewis, and B.~Li, ``Homotopic policy iteration-based learning
  design for unknown linear continuous-time systems,'' {\em Automatica},
  vol.~138, p.~110153, 2022.

\bibitem{10342780}
C.~Chen, F.~L. Lewis, K.~Xie, and S.~Xie, ``Adaptive optimal control of unknown
  nonlinear systems via homotopy-based policy iteration,'' {\em IEEE Trans.
  Autom. Control}, vol.~69, no.~5, pp.~3396--3403, 2024.

\bibitem{10887315}
Y.-S. Ma, J.~Sun, Y.~Xu, S.-S. Cui, and Z.-G. Wu, ``Adaptive dynamic
  programming for optimal control of unknown \emph{LTI} system via interval
  excitation,'' {\em IEEE Trans. Autom. Control}, vol.~70, no.~7,
  pp.~4896--4903, 2025.

\bibitem{PANG2025112227}
Z.~Pang, S.~Tang, J.~Cheng, and S.~He, ``Scaling policy iteration based
  reinforcement learning for unknown discrete-time linear systems,'' {\em
  Automatica}, vol.~176, p.~112227, 2025.

\bibitem{li2024data}
D.~Li and J.~Dong, ``Data-based efficient off-policy stabilizing optimal
  control algorithms for discrete-time linear systems via damping
  coefficients,'' {\em arXiv preprint arXiv:2412.20845}, 2024.

\end{thebibliography}
\end{document}